\documentstyle[prb,aps,floats,psfig,oldlfont]{revtex}

\begin{document}
\hsize\textwidth\columnwidth\hsize\csname@twocolumnfalse\endcsname

\title{Dynamics of a mesoscopic qubit under continuous quantum measurement}
\author{Hsi-Sheng Goan \cite{goan} and Gerard J.~Milburn}
\address{Center for Quantum Computer Technology and Department of
Physics, University of Queensland,
Brisbane, Qld 4072 Australia}
\maketitle
\draft

\begin{abstract}
We present the conditional quantum dynamics of an electron tunneling
between two quantum dots subject to a measurement using a low transparency
point contact or tunnel junction. 
The double dot system forms a single qubit and the
measurement corresponds to a continuous in time readout of the occupancy of
the quantum dot. 
We illustrate the difference between conditional and
unconditional dynamics of the qubit. 
The conditional dynamics is discussed in two
regimes depending on the rate of tunneling through the point contact:
quantum jumps, in which individual electron tunneling current events can be
distinguished, and a diffusive dynamics in which individual events 
are ignored, and the time-averaged current is considered as a continuous 
diffusive variable. 
We include the effect of inefficient measurement and the influence of the
relative phase between the two tunneling amplitudes of the double
dot/point contact system.
\end{abstract}



\section{Introduction}

In condensed matter physics, measurements are usually made on
many identical quantum systems
prepared at the same time.
For example, in nuclear or electron magnetic resonance experiments, 
an ensemble of systems of nuclei and 
electrons are probed to obtain the resonance signals. 
In this case, the measurement result is 
an average response of the ensemble. 
On the other hand,  various proposed condensed-matter quantum computer
architectures \cite{Kane98,Loss98,Schoen97,Bonadeo98} demand
the need to readout physical properties of a single electronic qubit, 
such as charge or spin at a single electron level. 
This is a non-trivial problem since it involves 
an individual quantum particle measured by a practical detector 
in a realistic environment. It is particularly important to take account 
of the decoherence introduced by the measurements on the qubit 
as well as to understand how the quantum state of the qubit, 
conditioned on a particular single realization of measurement, 
evolves in time for the purpose of quantum computing. 

We consider, in this paper, the
problem of an electron tunneling between two coupled quantum dots
(CQD's), a two-state quantum system (qubit),
using a low-transparency point contact (PC) or tunnel junction
as a detector (environment) continuously measuring the position
of the electron, schematically illustrated in Fig.\ \ref{fig:PC}. 
This problem has been extensively studied in Refs.\
\cite{Gurvitz97,Gurvitz98,Korotkov99,Korotkov99b,Korotkov00,Makhlin98,Makhlin00,Korotkov00b,Korotkov00c,Goan00}.
The case of measurements by a general quantum point contact detector 
with arbitrary transparency
has also been investigated in Refs.\
\cite{Aleiner97,Levinson97,Stodolsky98,Buttiker00,Hackenbroich98,Averin00}.
In addition, a similar system measured by a single electron transistor
rather than a PC has been studied
in Refs.\ \cite{Shnirman98,Makhlin98,Korotkov99b,Makhlin00,Korotkov00c,Averin00b,Brink00}.
The influence of the detector (environment) on the measured system can
be determined by the reduced density matrix obtained by tracing out
the environmental degrees of the freedom in the total, system plus
environment, density matrix. 
The master equation (or rate equations
for the reduced-density-matrix elements) for the CQD system (qubit) 
has been
derived and analyzed 
in Refs.\ \cite{Gurvitz97,Goan00}. 
However, this (unconditional) master equation 
describes only the ensemble average property
for the CQD system. 
Integrating or tracing out
the PC environmental degrees of the freedom is equivalent to 
completely ignoring or averaging over the results of all
measurement records (electron current records in this case).
In this sense, the PC 
is treated as a pure environment rather than a measurement device
which can provide information about the change of the state of the qubit.
On the other hand, if a measurement
is made on the system and the results are available, the state or density
matrix is a
conditional state conditioned on the measurement results.
Hence the deterministic, unconditional master equation cannot
describe the conditional dynamics of the qubit in a single realization
of continuous measurements which reflects the stochastic nature of an
electron tunneling through the PC barrier.

Korotkov \cite{Korotkov99,Korotkov00,Korotkov00c} 
has obtained the Langevin rate
equations for the CQD system measured by a ideal PC detector.
These rate equations describe the random
evolution of the density matrix that both conditions, and is conditioned
by, the
PC detector output.
Recently, we \cite{Goan00}
presented a {\em quantum trajectory}
\cite{Carmichael93,Dalibard92,Gisin92,Wiseman93,Gagen93,Hegerfeldt93,Wiseman96,Presilla96,Mensky98,Plenio98,Gardiner00}
measurement analysis of the same system.
We found that the conditional dynamics of the 
CQD system can be described by the
stochastic Schr\"{o}dinger equation
for the conditioned state vector, provided that the information
carried away from the CQD system by the PC reservoirs can be recovered by
the perfect detection of the measurements.
We also analyzed the localization rates at which the
qubit becomes localized in one of the two states 
when the coupling frequency $\Omega$ between the states is zero.
We showed that the localization time discussed there
is slightly different from
the measurement time defined in
Refs.\ \cite{Shnirman98,Makhlin98,Makhlin00}.
The mixing rate at which the two possible states of
the qubit become mixed when $\Omega\neq 0$ was calculated as well
and found in agreement with
the result in Ref.\ \cite{Makhlin98,Makhlin00}.
In this paper,
we focus on the qubit dynamics conditioned on a particular realization
of the actual measured current through the PC device.
Especially, we take into account the effect of inefficient measurement
on the conditional dynamics and 
illustrate the conditional quantum 
evolutions by numerical simulations.

The problem for a ``non-ideal'' detector was discussed 
in Refs.\ \cite{Korotkov99,Korotkov99b,Korotkov00}.  
There the non-ideality of the detector
is modeled as two ideal detectors ``in parallel'' with inaccessible
output of the second detector. The information loss is due to the
interaction with the second detector, treated as a ``pure
environment'', (which does not affect the detector current). As a
consequence, the decoherence rate, $\Gamma_{tot}$, in that case is larger
than the decoherence rate for the PC as an environment
alone, $\Gamma_{tot}-\Gamma_d=\gamma_d>0$. Hence an extra decoherence term,
$-\gamma_d \rho_{ab}$, for example, is added in the rate equation
$\dot{\rho}_{ab}$. However, this approach does not account for the
inefficiency in the measurements, which arises when the detector
sometimes misses detection. In that case, there is still only one PC
detector (environment) and disregarding all measurement records leads
to $\Gamma_{tot}=\Gamma_d$. Furthermore, the detector current is
affected and in fact reduced by the inefficiency in the
measurements. 

In this paper, we analyze the conditional qubit dynamics analytically
and numerically.
We take into account the effect of inefficient measurement
of the PC detector on the dynamics of the qubit.
The different behavior of unconditional and
conditional evolution is demonstrated.
We present the conditional quantum dynamics over the full range of 
behavior, from quantum jumps to quantum diffusion. \cite{Goan00}
In Refs.\ \cite{Gurvitz97,Korotkov99,Korotkov00,Korotkov00c}, the two
tunneling amplitudes of the CQD-PC model were assumed to be real.
In Ref.\ \cite{Goan00}, the relative phase between them 
was taken into account.
Here, we discuss and illustrate furthermore their influence 
on the qubit dynamics.
In Sec.~\ref{sec:masterEq}, we describe the model Hamiltonian and the
unconditional master equation.
We then obtain in Sec.~\ref{sec:conditional}
the quantum-jump and quantum-diffusive, 
conditional master equations 
for the case of inefficient measurements.
Sec.~\ref{sec:dynamics} is devoted to
the analysis for the qubit dynamics. 
Numerical simulations of the conditional evolution are presented
in this section.
Finally, a short
conclusion is given in Sec.~\ref{sec:conclusion}. 
In the Appendix, 
the stationary noise power spectrum of
the current fluctuations through the PC barrier
is calculated in terms of the quantum-jump formalism.

\section{Unconditional master equation for the CQD and PC model}
\label{sec:masterEq}

\begin{figure}
\centerline{\psfig{file=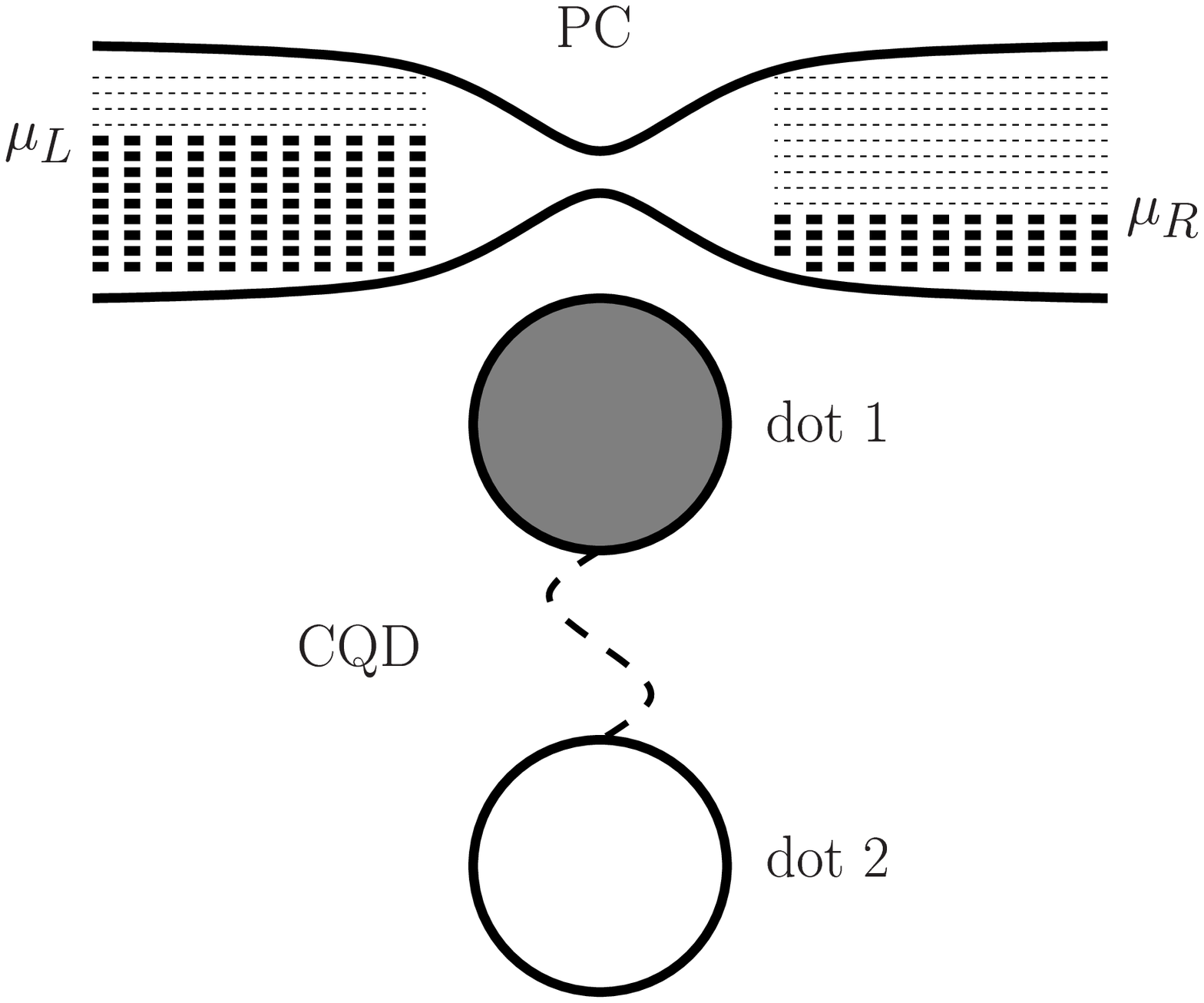,width=\linewidth,angle=0}}
\caption{Schematic representation of 
an electron tunneling between two coupled quantum dots
(CQDs), a two-state quantum system (qubit),
using a low-transparency point contact (PC) or tunnel junction
as a detector (environment)
continuously measuring the position
of the electron.
Here $\mu_L$ and $\mu_R$ stand for the
chemical potentials in the left and right
reservoirs respectively.}
\label{fig:PC}
\end{figure}

The appropriate way to approach quantum measurement problems may
be to treat the measured system, the detector (environment), and the
coupling between them together microscopically. 
Following from Refs.\ \cite{Gurvitz97,Korotkov99,Korotkov00}, 
We describe the whole system (see Fig.\ \ref{fig:PC}) 
by the following Hamiltonian: 
\begin{equation}
{\cal H}= {\cal H}_{CQD}+{\cal H}_{PC}+{\cal H}_{OSC}+{\cal H}_{coup}
\end{equation}
where
\begin{eqnarray}
{\cal H}_{CQD}&=&\hbar\left[ \omega_1 c_1^\dagger c_1
+\omega_2c_2^\dagger c_2
+\Omega(c_1^\dagger c_2+ c_2^\dagger c_1)\right],
\label{HCQD} \\
{\cal H}_{PC}&=&\hbar \sum_k
\left(\omega_k^L a_{Lk}^\dagger a_{Lk}
+\omega_k^R a_{Rk}^\dagger a_{Rk}\right)
+ \sum_{k,q}
\left(T_{kq}a_{Lk}^\dagger a_{Rq}+ T^*_{qk}a_{Rq}^\dagger a_{Lk} \right),
\label{HPC} \\
{\cal H}_{coup}&=&\sum_{k,q} c_1^\dagger c_1
\left(\chi_{kq}a_{Lk}^\dagger a_{Rq}
+ \chi^*_{qk}a_{Rq}^\dagger a_{Lk} \right).
\label{Hcoup}
\end{eqnarray}
${\cal H}_{CQD}$ represents the effective tunneling Hamiltonian
for the measured CQD system (mesoscopic qubit).
For simplicity, we assume strong inner and inter dot Coulomb
repulsion, so only one electron can occupy this CQD system. We label
each dot with an index $1,2$ (see Fig.\ \ref{fig:PC})
and let $c_i$  ($c_i^\dagger$) and
$\hbar\omega_i$ represent the electron annihilation (creation)
operator and energy for a single electron state in each dot
respectively. The coupling between these two dots is given by
$\hbar\Omega$. The tunneling Hamiltonian for the PC
detector is represented by ${\cal H}_{PC}$ where $a_{Lk}$, $a_{Rk}$
and $\hbar\omega_k^L$, $\hbar\omega_k^R$ are respectively the fermion
(electron) field annihilation operators and energies for the left and
right reservoir states at wave number $k$. One should not be confused by
the electron in the CQD with the electrons in the PC reservoirs. The
tunneling matrix element between states $k$ and $q$ in left and right
reservoir respectively is given by $T_{kq}$. 
Eq. (\ref{Hcoup}), ${\cal H}_{coup}$, 
describes the interaction between the detector and the
measured system, depending on which dot is occupied. When the electron
in the CQD system is near the PC (i.e., dot $1$ is occupied),
there is a change in the PC tunneling barrier. This barrier change
results in a
change of the effective tunneling amplitude from $T_{kq}\rightarrow
T_{kq}+\chi_{kq}$. As a consequence, the current through the PC is
also modified. This changed current can be detected,
and thus a measurement of the location of the electron in the CQD system is
effected.

The (unconditional) master equation of the reduced density matrix  
for the CQD system (qubit) 
has been obtained in Refs. \cite{Gurvitz97,Goan00}. 
Under similar assumptions and approximations as in Ref.\ \cite{Goan00}, the 
zero-temperature, \cite{temperature} 
Markovian master equation of the qubit can be written as:
\begin{mathletters}
\label{masterEquation}
\begin{eqnarray}
\dot{\rho}(t)&=&-\frac{i}{\hbar}[{\cal H}_{CQD}, \rho(t)]
+{\cal D}[{\cal T}+{\cal X} n_1]\rho(t)
\label{masterEq}
\\
&\equiv&{\cal L} \rho(t),
\label{Liouvillian}
\end{eqnarray}
\end{mathletters}
where $n_1=c_1^\dagger c_1$ is the 
occupation number operator for dot 1 and 
the parameters ${\cal T}$ and ${\cal X}$
are given by
\begin{mathletters}
\label{rate}
\begin{eqnarray}
|{\cal T}|^2&=&D= 2\pi e |T_{00}|^2 g_L g_R V/\hbar,
\label{D}
\\
|{\cal T}+{\cal X}|^2&=&D'
= 2\pi e |T_{00}+\chi_{00}|^2 g_L g_R V/\hbar,
\label{D'}
\end{eqnarray}
\end{mathletters}
where $D$ and $D'$ are the average electron tunneling rates
through the PC barrier 
without and with the presence of the electron in dot 1 respectively.
In Eq.\ (\ref{rate}), 
$eV=\mu_L-\mu_R$ is the external bias applied across the PC
(where $e$ is the electronic charge, and 
$\mu_L$ and $\mu_R$ stand for the chemical potentials in the
left and right reservoirs respectively),
$T_{00}$ and $\chi_{00}$ are energy-independent tunneling amplitudes
near the average chemical potential, and $g_L$ and $g_R$ are the
energy-independent density of states for the left and right reservoirs.
In Eq.\ (\ref{masterEq}), the superoperator
\cite{Wiseman93,Wiseman93b,Wiseman96} ${\cal D}$
is defined as: 
\begin{equation}
{\cal D}[B]\rho={\cal J}[B]\rho - {\cal A}[B]\rho,
\label{defcalD}
\end{equation}
where
\begin{eqnarray}
{\cal J}[B]\rho &=& B \rho B^\dagger,
\label{defcalJ}\\
{\cal A}[B]\rho &=& (B^\dagger B \rho +\rho B^\dagger B)/2.
\label{defcalA}
\end{eqnarray}
Finally, Eq.\ (\ref{Liouvillian}) defines the Liouvillian operator ${\cal L}$.
The form of the master equation (\ref{masterEq}), defined through
the superoperator ${\cal D}[B]\rho(t)$, preserves the positivity of the
density matrix operator $\rho(t)$. Such a Markovian master equation
is called a Lindblad \cite{Lindblad73} form.

Evaluating the density matrix operator
in the same basis as in Ref.\ \cite{Gurvitz97}, we obtain
\begin{mathletters}
\label{rateEq}
\begin{eqnarray}
\dot\rho_{aa}(t)& = &i\Omega[\rho_{ab}(t)-\rho_{ba}(t)]\;,
\label{rateEqa}\\
\dot\rho_{ab}(t)& = & i{\cal E}\rho_{ab}(t)+i\Omega[\rho_{aa}(t)
-\rho_{bb}(t)]-(|{\cal X}|^2/2)\rho_{ab}(t)
+i\, {\rm Im} ({\cal T}^* {\cal X})\rho_{ab}(t),
\label{rateEqc}
\end{eqnarray}
\end{mathletters}
where $\hbar {\cal E}=\hbar (\omega_2-\omega_1)$ is the energy
mismatch between the two dots,
$\rho_{ij}(t)=\langle i|\rho(t)| j\rangle$,
and $\rho_{aa}(t)$ and $\rho_{bb}(t)$ are the probabilities
of finding the electron in dot 1 and dot 2 respectively.
The corresponding logical qubit states 
are then $|a\rangle$ and $|b\rangle$.  
The rate equations for the other two density matrix elements can be
easily obtained from the relations: $\rho_{bb}(t)=1-\rho_{aa}(t)$ and
$\rho_{ba}(t)=\rho_{ab}^*(t)$.
Compared to an isolated CQD system, the presence of
the PC detector (environment) introduces two effects to the CQD system.
First,  the imaginary part of 
$({\cal T}^* {\cal X})$
(the last term in Eq.\ (\ref{rateEqc})) causes an effective
shift in the
energy mismatch between the two dots.
Second, it generates a decoherence
(dephasing) rate
\begin{equation}
\Gamma_d={|{\cal X}|^2}/{2}
\label{dephaserate}
\end{equation}
for the off-diagonal density matrix elements.
The relative phase between the two complex tunneling
amplitudes may produce additional effects on
conditional dynamics of the CQD system as well.
This will be shown later when we discuss conditional dynamics.
Physically, the presence of the electron in dot 1 (state  $|a\rangle$)
raises the effective tunneling barrier of the PC
due to electrostatic repulsion.
As a consequence,
the effective tunneling amplitude becomes lower, i.e.,
$D'=|{\cal T}+{\cal X}|^2<D=|{\cal T}|^2$.
This sets a condition on the relative phase $\theta$
between ${\cal X}$ and ${\cal T}$:
$\cos\theta<-|{\cal X}|/(2|{\cal T}|)$.

\section{Conditional master equation}
\label{sec:conditional}
Equation (\ref{masterEquation}) describes the time evolution of 
reduced density matrix
when all the measurement results are ignored, or averaged over.
To make contact with a single realization of the measurement records
and study the stochastic evolution of the quantum state, conditioned on a
particular measurement realization, we 
obtain in this section the
conditional master equation.

The measurable quantities, such as accumulated number of
electrons tunneling through the PC barrier, are stochastic.
On average of course the same current flows in
both reservoirs. However, the current is actually made up of
contributions from random pulses in each reservoir,
separated in time by the times at which the electrons tunnel through
the PC, assuming the response of the reservoirs is fast. 
In this section, we first treat the electron tunneling current consisting of
a sequence of random $\delta$ function pulses. In other words, the measured
current is regarded as a series of point processes
(a quantum-jump model) \cite{Wiseman93,Plenio98}.
The case of quantum diffusion will be analyzed later.

For almost-all infinitesimal time intervals, the
measurement result is null 
(no detection of an electron tunneling through the PC barrier). 
The system in this
case changes infinitesimally, but not unitarily. The nonunitary component
reflects the
changing probabilities for future events conditioned on past null events.
At randomly
determined times (conditionally Poisson distributed), there is a
detection result. When this occurs, the system undergoes a finite
evolution, called a {\em quantum jump}.

Formally, we can write the current through the PC as
\begin{equation}
i(t) = e\, {dN(t)}/{dt},
\label{current}
\end{equation}
where $dN(t)$ is a classical point
process which represents the number (either zero or one) of tunneling
events seen in an infinitesimal time $dt$.
We can think of $dN(t)$ as the increment in the accumulated
number \cite{Gurvitz97,Shnirman98,Makhlin00} of electrons $N(t)$
in the drain in time $dt$.
The point process is formally defined \cite{Goan00} by the
conditions on the classical random variable $dN_c(t)$:
\begin{mathletters}
\label{pointprocess}
\begin{eqnarray}
[dN_c(t)]^{2} &=& dN_c(t),  \label{dN} \\
E[dN_c(t)] &=& \zeta\, {\rm Tr}[\tilde{\rho}_{1c} (t+dt)]
=\zeta\,  [D+(D'-D)\langle n_1\rangle_c(t)] dt.
\label{ineffdNav}
\end{eqnarray}
\end{mathletters}
where $\langle n_1\rangle_c(t)={\rm Tr}[n_1\rho_c(t)]$, 
$E[Y]$ denotes an ensemble
average of a classical stochastic process $Y$, and
\begin{equation}
\tilde\rho_{1c}(t+dt)={\cal J}[{\cal T}+{\cal X} n_1]\rho_c(t){dt}
\label{measuretilderho1}
\end{equation}
is the unnormalized density matrix \cite{Goan00} given the result of
an electron tunneling through the PC barrier
at the end of the time interval $[t,t+dt)$. 
We explicitly use the subscript $c$ to indicate that the quantity
to which it is attached is conditioned on previous measurement
results, the occurrences (detection records) of
the electrons tunneling
through the PC barrier in the past. 
Note that the density matrix
$\rho_c(t)$ is not the solution of the unconditional reduced master
equation, Eq.\ (\ref{masterEq}). It is actually conditioned
by $dN_c(t')$ for $t'<t$.
Equation (\ref{dN}) simply
states that $dN_c(t)$ equals either zero or one, which is why it is
called a point process. 
Equation (\ref{ineffdNav}) indicates that the ensemble
average of $dN_c(t)$ equals the probability (quantum average) of
detecting electrons tunneling through the PC barrier in time $dt$.
The inefficiency in the measurements, which arises when the detector
sometimes misses detection, is taken into account here. 
The factor $\zeta \leq 1$ represents 
the fraction of detections which
are actually registered by the PC detector.
The value $\zeta=1$ then corresponds to 
a perfect detector or efficient measurement.
By using Eq.\ (\ref{current}), equation (\ref{ineffdNav}) 
with $\zeta=1$ states
that the average current is $eD$ when dot 1 is empty, and is
$eD'$ when dot 1 is occupied.
In Ref.\ \cite{Korotkov00c} the case of 
inefficient measurements is discussed
in terms of insufficiently small readout period. 
In other words, the bandwidth of the
measurement device is not large enough to 
resolve and record every electron tunneling through the 
PC barrier.

By following the similar derivation as in Ref.\ \cite{Goan00}, 
the stochastic master equation of the density matrix operator, 
conditioned on the observed event in the efficient
measurement in time $dt$ can be obtained:
\begin{eqnarray}
d\rho_c(t)
&=&dN_c(t)\left [\frac{{\cal J}[{\cal T}+{\cal X} n_1]}
{{\cal P}_{1c}(t)}
-1\right ]\rho_c(t)
\nonumber \\
&& +\, dt \left\{
-{\cal A}[{\cal T}+{\cal X} n_1]\rho_c(t)
+(1-\zeta){\cal J}[{\cal T}+{\cal X} n_1] \rho_c(t)
+\zeta \,  {\cal P}_{1c}(t) \rho_c(t)
+\frac{i}{\hbar}[{\cal H}_{CQD},\rho_c(t)] \right\},
\label{condmasterEq}
\end{eqnarray}
where
\begin{equation}
{\cal P}_{1c}(t)= D+(D'-D)\langle n_1\rangle_c(t).
\end{equation}
Averaging this equation over the observed stochastic process, by
setting $E[dN_c(t)]$ equal to its expected value Eq.\ (\ref{ineffdNav}),
gives the unconditional, deterministic
master equation (\ref{masterEq}).

Similarly, the extension to the case of 
quantum diffusion can be carried out as in Ref.\ \cite{Goan00}.
In this case, the
average electron tunneling rate is very large compared to the extra change
of the tunneling rate due to the presence of the electron in the
dot closer to the PC.
In addition, individual electrons tunneling through the PC are ignored
and  time averaging  of the currents is performed.
This allows electron counts, or accumulated electron number, to be
considered as a continuous diffusive variable satisfying a
Gaussian white noise distribution \cite{Goan00,Wiseman93b}
in time $\delta t$:
\begin{equation}
\delta N(t)=\{\zeta \,  |{\cal T}|^2 [1+2\, \epsilon\, 
\cos\theta\,\langle n_1\rangle_c(t)]
+\sqrt{\zeta\, } \, |{\cal T}| \xi(t)\} \delta t,
\label{deltaN}
\end{equation}
where $\epsilon=(|{\cal X}|/|{\cal T}|)\ll 1$, 
$\theta$ is the relative phase between ${\cal X}$ and ${\cal T}$,
and $\xi(t)$ is a Gaussian white noise characterized by
\begin{equation}
E[\xi(t)]=0, \quad E[\xi(t)\xi(t')]=\delta(t-t').
\label{xi}
\end{equation}
Here $E$ denotes an ensemble average and $\delta(t-t')$ is a delta function.
In stochastic calculus \cite{Gardiner85,Oksendal92},
$\xi(t)dt=dW(t)$ is known as the infinitesimal
Wiener increment with the property $E[(\xi(t)dt)^2]=E[dW(t)^2]=dt$.
In obtaining Eq.\ (\ref{deltaN}), we have assumed that 
$2|{\cal T}||{\cal X}|\, \cos\theta \gg |{\cal X}|^2$.
Hence, for the quantum-diffusive equations obtained later, we should
regard, to the order of magnitude, that 
$|\cos\theta|\sim O(1)\gg \epsilon=(|{\cal X}|/|{\cal T}|)$
and $|\sin\theta|\sim O(\epsilon)\ll 1$.
The quantum-diffusive conditional master equation can be found as:
\begin{eqnarray}
\dot{\rho}_c(t)
&=&
-\frac{i}{\hbar}[{\cal H}_{CQD}, \rho_c(t)]
+{\cal D}[{\cal T}+{\cal X}n_1]\rho_c(t)
\nonumber \\
&&
+\xi(t)\frac{\sqrt{\zeta\, } \, }{|{\cal T}|}
\left[{\cal T}^* {\cal X}\,n_1\rho_c(t)+ {\cal X}^* {\cal T}\rho_c(t)n_1
-2\,{\rm Re}({\cal T}^* {\cal X})\langle n_1\rangle_c(t)\rho_c(t)\right].
\label{diffusivemasterEq}
\end{eqnarray}
In arriving at Eq.\ (\ref{diffusivemasterEq}),
we have used the stochastic It\^{o} calculus 
\cite{Gardiner85,Oksendal92} for 
the definition of derivative as
$\dot{\rho}(t)
=\lim_{dt\rightarrow 0}[{\rho(t+dt)-\rho(t)}]/{dt}$.
This is in contrast to the definition,
$\dot{\rho}(t)
=\lim_{dt\rightarrow 0}[{\rho(t+dt/2)-\rho(t-dt/2)}]/{dt}$,
used in another stochastic calculus, the Stratonovich 
calculus \cite{Gardiner85,Oksendal92}.
It is easy to see that the ensemble
average evolution of Eq.\ (\ref{diffusivemasterEq}) 
reproduces the unconditional master equation
(\ref{masterEq}) by simply eliminating the white noise term using
Eq.\ (\ref{xi}).

If and only if detections are perfect (efficient measurement), 
i.e. $\zeta=1$, are the stochastic master equations 
for the conditioned density matrix operators, (\ref{condmasterEq}) 
and (\ref{diffusivemasterEq}), equivalent to the stochastic 
Schr\"{o}dinger equations 
(Eqs. (35) and (41) of Ref.\ \cite{Goan00}, respectively) 
for the conditioned states.
That is because 
in order for the system to be continuously described by a state
vector (rather than a general density matrix), it is necessary
(and sufficient) to have maximal knowledge of its change of state.
This requires perfect detection or efficient measurement, which
recovers and contains all the information lost from the system
to the reservoirs.
If the detection is not perfect and some information about the
system is {\em unrecoverable}, the evolution of the system can no longer
be described by a pure state vector. For the extreme case of zero
efficiency detection, the information
(measurement results at the detector) carried
away from the system to the reservoirs is (are) completely
ignored, so that the stochastic master equations
(\ref{condmasterEq}) and (\ref{diffusivemasterEq})
after being averaged over all possible
measurement records
reduces to the unconditional, deterministic master equation
(\ref{masterEq}), leading to decoherence for the system.

To make the quantum-diffusive, conditional stochastic
master equation (\ref{diffusivemasterEq})
more transparent, we
evaluate Eq.\ (\ref{diffusivemasterEq})
in the same basis as for Eq.\ (\ref{rateEq}) and
obtain:
\begin{mathletters}
\label{diffrateEq}
\begin{eqnarray}
\dot\rho_{aa}(t)& = &i\Omega[\rho_{ab}(t)-\rho_{ba}(t)]
+\sqrt{8\, \zeta \,  \Gamma_d} \, 
\cos\theta\, \rho_{aa}(t) \rho_{bb}(t)\xi(t)\;,
\label{diffrateEqa}\\
\dot\rho_{ab}(t)& = & i({\cal E}+ |{\cal T}||{\cal X}|\sin\theta)
\rho_{ab}(t)+i\Omega[\rho_{aa}(t)
-\rho_{bb}(t)]-\Gamma_d \, \rho_{ab}(t)
\nonumber \\
&&+\sqrt{2\, \zeta \, \Gamma_d}
\{\cos\theta\, [\rho_{bb}(t)-\rho_{aa}(t)]+i\sin\theta\} 
\, \rho_{ab}(t)\xi(t)\;,
\label{diffrateEqc}
\end{eqnarray}
\end{mathletters}
where we have set $|{\cal X}|=\sqrt{2 \Gamma_d}$.
Again, the ensemble average of Eq.\ (\ref{diffrateEq})
by eliminating the white noise terms
reduces to Eq.\ (\ref{rateEq}).
It is also easy to verify that for zero efficiency $\zeta=0$,
the conditional equations (\ref{condmasterEq}), (\ref{diffusivemasterEq}),
and (\ref{diffrateEq}), 
reduce to the corresponding 
unconditional ones, (\ref{masterEq}) and (\ref{rateEq}) respectively.
That is, the effect of averaging over all possible
measurement records is equivalent to the effect of completely ignoring
the detection records
or the effect of no detection results being available.

\section{Conditional dynamics under continuous measurements}
\label{sec:dynamics}

In this section, we analyze the qubit dynamics in detail and
present the numerical simulations for the time evolution.
We represent the qubit density matrix elements
in terms of Bloch sphere variables since some
physical insights into the dynamics of the qubit can sometimes be
more easily visualized in this representation. Denoting 
\begin{eqnarray}
I&=& c^\dagger_{2}c_{2}+c^\dagger_{1}c_{1},\\
\sigma_x & = & c^\dagger_{2}c_{1}+c^\dagger_{1}c_{2}, \\
\sigma_y & = & -ic^\dagger_{2}c_{1}+ic^\dagger_{1}c_{2}, \\
\sigma_z & = & c^\dagger_{2}c_{2}-c^\dagger_{1}c_{1},
\end{eqnarray}
we write the density matrix operator for the CQD system (qubit)
in terms of the Bloch sphere
vector $(x,y,z)$ as:
\begin{mathletters}
\label{Blochsphere}
\begin{eqnarray}
\rho(t)&=&[I+x(t)\sigma_x
+y(t)\sigma_y+z(t)\sigma_z]/2
\label{Blochrho}
\\
&=&\frac{1}{2}\left (\begin{array}{cc}
1+z(t)&x(t)-iy(t)\\
x(t)+iy(t)& 1-z(t)\end{array}\right ).
\label{bloch}
\end{eqnarray}
\end{mathletters}
It is easy to see that ${\rm Tr}\rho(t)=1$, $I$ is a unit operator, 
$\sigma_i$ defined above satisfies the
properties of Pauli matrices, and
the averages of the
operators $\sigma_x$, $\sigma_y$, $\sigma_z$ are $x(t)$, $y(t)$, $z(t)$
respectively.
In this representation, the variable $z(t)$ represents the
population difference between the two dots. Especially, $z(t)=1$ and
$z(t)=-1$ indicate that the electron is localized in dot 2 and
dot 1 respectively. The value $z(t)=0$ corresponds to an equal
probability for the electron to be in each dot.
Generally the product of the
off-diagonal elements of $\rho(t)$ is smaller than the product of the
diagonal elements, leading to the relation $x^2(t)+y^2(t)+z^2(t)\leq
1$. When $\rho(t)$ is represented by a pure state, the equal sign
holds. In this case, the system state can be characterized by a point
$(x,y,z)$ on the Bloch unit sphere.

The master equations (\ref{masterEq}), (\ref{diffusivemasterEq})
and (\ref{condmasterEq}),
can be written as a set of coupled stochastic
differential equations in terms of the Bloch sphere variables.
By substituting Eq.\ (\ref{Blochrho}) into Eq.\ (\ref{masterEq}),
and collecting and equating the coefficients in front of $\sigma_x$,
$\sigma_y$, $\sigma_z$ respectively, the unconditional master equation
under the assumption of real tunneling amplitudes
is equivalent to the following equations:
\begin{mathletters}
\label{uncondBlochEq}
\begin{eqnarray}
\frac{dx(t)}{dt} &=&-({\cal E}+ |{\cal T}||{\cal X}|\sin\theta)
\, y(t)-\Gamma_d \, x(t), \\
\frac{dy(t)}{dt} &=&({\cal E}+ |{\cal T}||{\cal X}|\sin\theta)
\, x(t) -2\Omega \, z(t)
-\Gamma_d \, y(t) ,  \\
\frac{dz(t)}{dt} &=&2\Omega \, y(t).
\label{uncondz}
\end{eqnarray}
\end{mathletters}
Similarly for the quantum-diffusive, conditional
master equation (\ref{diffusivemasterEq}), we obtain
\begin{mathletters}
\label{diffBlochEq}
\begin{eqnarray}
\frac{dx_c(t)}{dt} &=&-({\cal E}+ |{\cal T}||{\cal X}|\sin\theta) 
\, y_c(t)-\Gamma_d \, x_c(t)
+\sqrt{2\, \zeta\, \Gamma_d} [-\sin\theta\, y_c(t)+\cos\theta
\, z_c(t) x_c(t)] \xi(t), \\
\frac{dy_c(t)}{dt} &=& ({\cal E}+ |{\cal T}||{\cal X}|\sin\theta)
\, x_c(t) -2\Omega \, z_c(t)
-\Gamma_d \, y_c(t) +\sqrt{2\, \zeta\, \Gamma_d}[\sin\theta\, x_c(t)+\cos\theta
\, z_c(t) y_c(t)] \xi(t),  \\
\frac{dz_c(t)}{dt} &=&2\Omega \, y_c(t)
-\sqrt{2\, \zeta \, \Gamma_d}\,\cos\theta \left[1-z^2_c(t)\right] \xi(t).
\label{diffz}
\end{eqnarray}
\end{mathletters}
Again the $c$-subscript is to emphasize that these variables refer to
the conditional state. It is trivial to see that 
Eq.\ (\ref{diffBlochEq})
averaged over the white noise
reduces to Eq.\ (\ref{uncondBlochEq}), provided that
$E[x_c(t)]=x(t)$ as well as similar replacements are performed
for $y_c(t)$ and $z_c(t)$.
The analogous calculation can be carried out for the quantum-jump,
conditional master equation (\ref{condmasterEq}). We obtain
\begin{mathletters}
\label{condBlochEq}
\begin{eqnarray}
dx_c(t)&=&dt \left( -[{\cal E}+ (1-\zeta)|{\cal T}||{\cal X}|\sin\theta] 
\, y_c(t)-(1-\zeta)\Gamma_d \, x_c(t)
-\frac{\zeta \,  (D'-D)}{2} z_c(t) x_c(t)\right)
\nonumber \\
&&-dN_c(t)\left(
\frac{2 |{\cal T}||{\cal X}|\sin\theta\, y_c(t)
+[2\Gamma_d-(D'-D)z_c(t)]x_c(t)}{2D+(D'-D)[1-z_c(t)]}\right)\, ,
\\
dy_c(t) &=&dt\left( [{\cal E}+ (1-\zeta)|{\cal T}||{\cal X}|\sin\theta]  
\, x_c(t)-(1-\zeta)\Gamma_d\, y_c(t)
-2\Omega \, z_c(t)-\frac{\zeta \,  (D'-D)}{2} z_c(t) y_c(t)\right)
\nonumber \\
&&-dN_c(t)\left(
\frac{-2 |{\cal T}||{\cal X}|\sin\theta\, x_c(t)
+[2\Gamma_d-(D'-D)z_c(t)]y_c(t)}{2D+(D'-D)[1-z_c(t)]}\right) \, ,
\\
dz_c(t) &=&dt \left(2\Omega \, y_c(t)
+\frac{\zeta \,  (D'-D)}{2}\left[1-z_c^2(t)\right]\right)
-dN_c(t)\left(\frac{(D'-D)[1-z^2_c(t)]}{2D+(D'-D)[1-z_c(t)]}\right) \, .
\label{condz}
\end{eqnarray}
\end{mathletters}
As expected, by using Eq.\ (\ref{ineffdNav}), the ensemble average
of Eq.\ (\ref{condBlochEq}) also reduces to the unconditional
equation (\ref{uncondBlochEq}).
One can also observe that 
for $\zeta=0$, the conditional equations, (\ref{condBlochEq})
and (\ref{diffBlochEq}),
reduce to the unconditional equation (\ref{uncondBlochEq})
as well. 
Next, we analyze the conditional qubit dynamics.
Part of the results in Sec.\ \ref{sec:jump-diff} have been reported in 
Ref.\ \cite{Goan01}.

\subsection{From quantum jumps to quantum diffusion}
\label{sec:jump-diff}

\begin{figure}
\centerline{\psfig{file=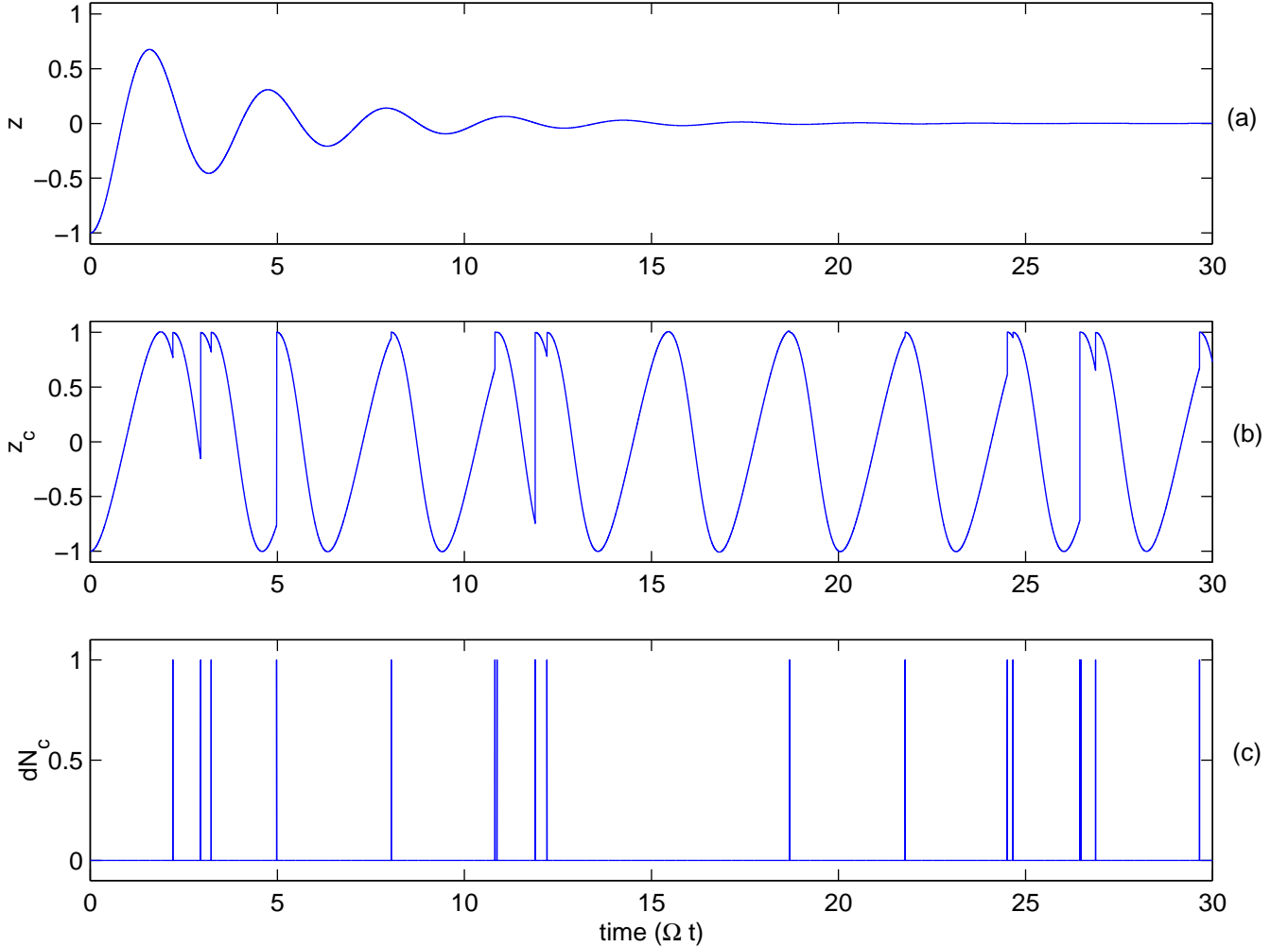,width=\linewidth,angle=0}}
\caption{Illustration for different behaviors between unconditional and
conditional evolutions. The initial qubit state is $|a\rangle$.
The parameters are: $\zeta=1$, 
${\cal E}=0$, $\theta=\pi$, $|{\cal T}|^2=|{\cal X}|^2=\Omega$,
and time is in unit of $\Omega^{-1}$.  
(a) unconditional, ensemble-averaged time evolution of $z(t)$,
which exhibits some oscillation and then approaches a zero steady
state value.
(b) conditional evolution of $z_c(t)$. The qubit starts an oscillation,
which is then interrupted by a quantum jump (corresponding to a 
detection of an electron passing through the PC barrier in (c)). 
After the jump, the
qubit state is reset to $|b\rangle$ and a new oscillation starts.
(c) randomly distributed moments of detections, which
correspond to the quantum jumps in (b).}
\label{fig:PC-jumps}
\end{figure}

Fig.\ \ref{fig:PC-jumps}(a)
shows the unconditional (ensemble average) time evolution of the
population difference $z(t)$ with the initial qubit state being in
state $|a\rangle$, i.e., dot 1 is occupied .  
The unconditional population difference $z(t)$, 
rises from $-1$, undergoing some oscillations, and then
tends towards zero, a steady (maximally mixed) state. 
On the other hand, the conditional
time evolution, conditioned on one possible individual realization of
the sequence of measurement results, behaves quite differently. 
We consider first the situation, where 
$D'=|{\cal T}+{\cal X}|^2=0$, discussed in Ref.\ \cite{Gurvitz98}. 
In this case,
due to the electrostatic repulsion generated by the electron, the PC
is blocked (no electron is transmitted) when dot 1 is
occupied. As a consequence, whenever there is a detection of an electron
tunneling through the PC barrier, the qubit state is collapsed into state
$|b\rangle$, i.e., dot 2 is occupied. 
The quantum-jump conditional evolution shown in Fig.\ \ref{fig:PC-jumps}(b) 
(using the same parameters and initial
condition as in Fig.\ \ref{fig:PC-jumps}(a)) 
is rather obviously
different from the unconditional one in Fig.\ \ref{fig:PC-jumps}(a).
The conditional time evolution is not smooth, but exhibits jumps, and it does
not tend towards a steady state.  One can see that initially the
system starts to undergo an oscillation. As the population
difference $z_c(t)$ changes in time, 
the probability for an electron
tunneling through the PC barrier increases. 
This oscillation is then interrupted
by the detection of an electron tunneling through the PC barrier,
which bring $z_c(t)$ to the value $1$, i.e., 
the qubit state is collapsed into state $|b\rangle$. 
Then the whole process starts again.
The randomly distributed moments of detections, $dN_c(t)$,
corresponding to the quantum jumps in Fig.\ \ref{fig:PC-jumps}(b) 
is illustrated in Fig.\ \ref{fig:PC-jumps}(c).
Although little similarity can be observed between
the time evolution in
Fig.\ \ref{fig:PC-jumps}(a) and (b), 
averaging over
many individual realizations shown in Fig.\ \ref{fig:PC-jumps}(b) 
leads to a closer and
closer approximation of the ensemble average 
in Fig.\ \ref{fig:PC-jumps}(a).

\begin{figure}
\centerline{\psfig{file=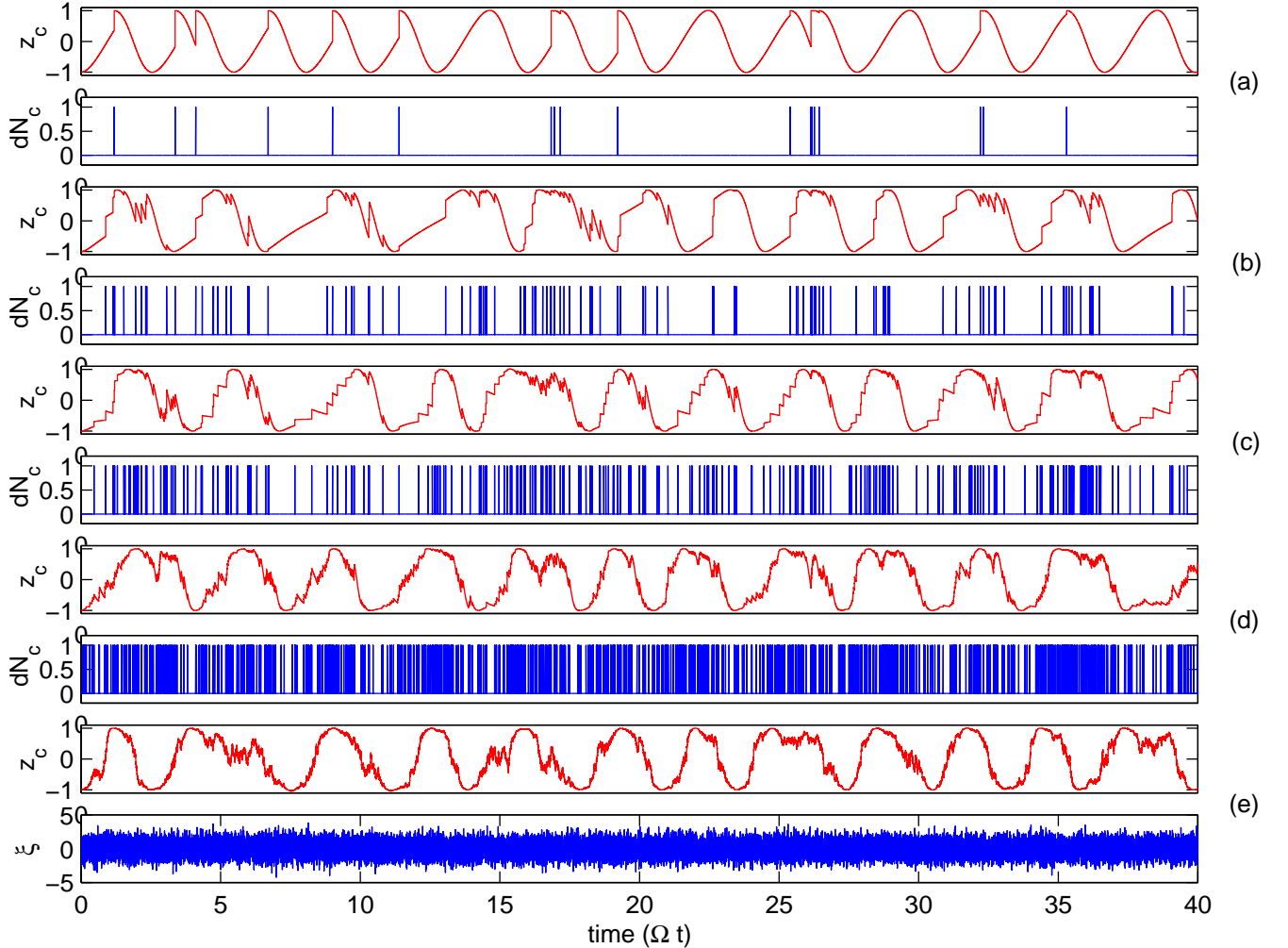,width=\linewidth,angle=0}}
\caption{Transition from quantum jumps to quantum diffusion.
The initial qubit state is $|a\rangle$.
The parameters are: $\zeta=1$, 
${\cal E}=0$, $\theta=\pi$, $|{\cal X}|^2=\Omega$,
and time is in unit of $\Omega^{-1}$.
(a)--(d) are the quantum-jump, conditional evolutions of $z_c(t)$ 
and corresponding detection moments 
with different $|{\cal T}|/|{\cal X}|$ ratios:
(a) 1, (b) 2, (c) 3, (d) 5. With increasing $|{\cal T}|/|{\cal X}|$
ratio, jumps become more frequent but smaller in amplitude.
(e) represents the conditional evolutions of $z_c(t)$ 
in the quantum diffusive limit.
The variable $\xi(t)$, appearing in the expression of current through PC in 
quantum-diffusive limit, 
is a Gaussian white noise with zero mean and unit variance.}
\label{fig:PC-jump-diff}
\end{figure}

Next we illustrate how the
transition from the quantum-jump picture to the quantum-diffusive
picture takes place.
In  Ref.\ \cite{Goan00} and Sec.\ \ref{sec:conditional}, 
we have seen that the quantum-diffusive
equations can be obtained from the quantum-jump description under the
assumption of $|{\cal T}| \gg |{\cal X}|$.  
In Fig.\ \ref{fig:PC-jump-diff}(a)--(d) 
we plot conditional, quantum-jump evolution of $z_c(t)$ and 
the corresponding moments of detections $dN_c(t)$,  
with different $(|{\cal T}|/|{\cal X}|)$ ratios.
Each jump (discontinuity) in the $z_c(t)$ curves 
corresponds to the detection
of an electron through the PC barrier.
One can clearly
observes that with increasing $(|{\cal T}|/|{\cal X}|)$ ratio, the
number of jumps increases.
The amplitudes of the jumps of $z_c(t)$, however, 
decreases from $D'=0$ with
the certainty of the qubit being in 
state $|b\rangle$ to the case of
$(D-D')<<(D+D')$ with a smaller probability of 
finding the qubit in state $|b\rangle$.
Nevertheless, the population
difference $z_c(t)$ always jumps up
since $D=|{\cal T}|^2>D'=|{\cal T}+{\cal  X}|^2$. 
In other words, whenever there is a detection of an
electron passing through PC, dot 2 is more likely occupied than dot 1.
The case for quantum diffusion using Eq.\ (\ref{diffBlochEq})
is plotted in Fig.\ \ref{fig:PC-jump-diff}(e).
In this case, very small jumps occur very frequently.
We can see that the behavior of $z_c(t)$
for $|{\cal T}|=5|{\cal X}|$ in the
quantum-jump case shown in Fig.\ \ref{fig:PC-jump-diff}(d) is already 
very close to that of quantum diffusion shown in 
Fig.\ \ref{fig:PC-jump-diff}(e).
To minimize the number of controllable variables,
the same randomness is applied to 
produce the quantum-jump,
conditional evolutions in Fig.\ \ref{fig:PC-jump-diff}(a)--(d). 
This, however, does not mean that they would have had the same
detection output, $dN_c(t)$. 
The number of tunneling events in time $dt$, $dN_c(t)$, does not
depend on the randomness alone. It also depends on 
$|{\cal T}|$, $|{\cal X}|$, and $\theta$, and has to satisfy 
Eq.\ (\ref{ineffdNav}) in a self-consistent manner. 
In fact, it both conditions and is conditioned by the conditional
qubit density matrix. 
Note that the unconditional evolution does not depend on the parameter
$|{\cal T}|$ when $\theta=\pi$ (see Eq.\ (\ref{uncondBlochEq})). 
This implies that depending on the actual measured detection events,
different measurement schemes (measurement devices with
different tunneling barriers or
different values of $|{\cal T}|$  when $\theta=\pi$) give different
conditional quantum evolutions. But they would have the same ensemble
average property if other parameters and the initial condition are the same.
Hence, averaging over all possible realizations, for each measurement scheme 
in Fig.\ \ref{fig:PC-jump-diff}, will lead to the same ensemble
average behavior shown in Fig.\ \ref{fig:PC-jumps}(a).

\subsection{Quantum Zeno effect}
\label{sec:Zeno}

\begin{figure}
\centerline{\psfig{file=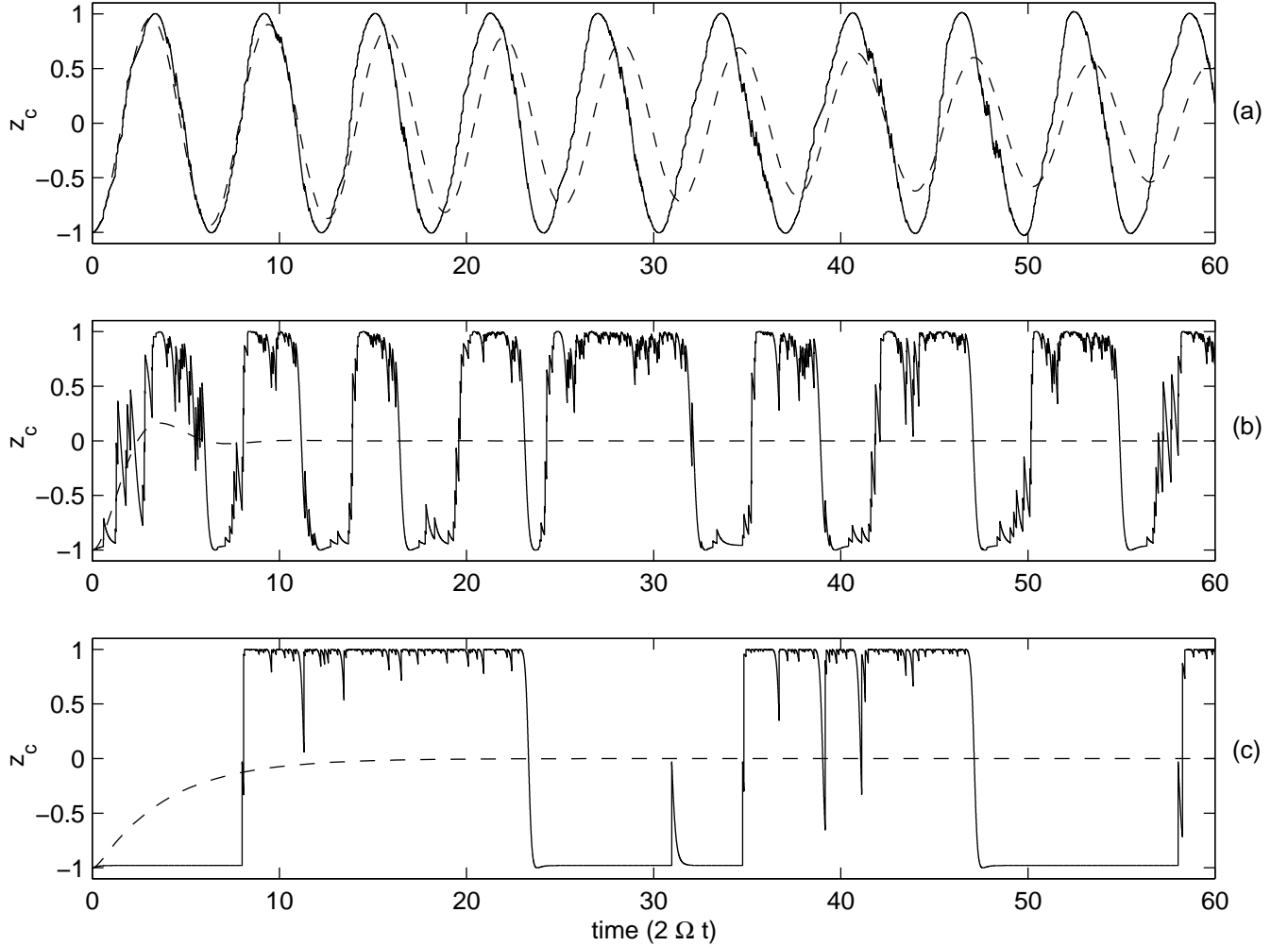,width=\linewidth,angle=0}}
\caption{Illustration of the quantum Zeno effect. 
Both conditional (in solid line) and unconditional (in dashed line)  
evolutions of the population difference 
for different ratios
of (a)$(\Gamma_d/\Omega)$=0.04, (b) 2, (c) 8, are shown.
The initial qubit state is $|a\rangle$.
The other parameters are: $\zeta=1$, 
${\cal E}=0$, $\theta=\pi$, $|{\cal T}|^2=20\Omega$,
and time is in unit of $(2\Omega)^{-1}$.
Increasing $(\Gamma_d/\Omega)$ ratio increases the
period of coherent oscillations 
between the qubit states, 
while the time of a transition (switching time)
decreases.} 
\label{fig:PC-Zeno}
\end{figure}

The quantum Zeno effect can be naturally described by the conditional
dynamics. The case for quantum diffusion has been discussed in 
Ref. \cite{Korotkov99,Korotkov00}. Here, for completeness, we discuss the
quantum-jump case. The quantum Zeno effect states that repeated
observations of the system slow down transitions between quantum
states due to the collapse of the wave function into the observed state. 
Alternatively, the interaction with one measurement apparatus destroys
the quantum coherence (oscillations) 
between $|a\rangle$ and $|b\rangle$ 
at a rate that is much faster than the tunneling rate $\Omega$. 
For fixed $\Omega$, $|\cal T|$ and $\theta$, 
by increasing the interaction with the PC detector 
$|{\cal X}|=\sqrt{2\Gamma_d}$, we increase the number and amplitude of
jumps and hence the probability of the 
wave function being collapsed to the localized state. The time
evolutions of the population difference $z_c(t)$ for different ratios
of $(\Gamma_d/\Omega)$ are shown in Fig.\ \ref{fig:PC-Zeno}. 
Here, the initial qubit state is $|a\rangle$, and other parameters are: 
$\zeta=1$, ${\cal E}=0$, $\theta=\pi$, $|{\cal T}|^2=10\Omega$.
We can observe that the 
period of coherent oscillations between the two qubit states
increases with increasing $(\Gamma_d/\Omega)$, 
while the time of a transition (switching time) decreases.
In the limit of vanishing $\Omega$, a transition from one qubit state
to the other state takes a time (switching time) 
of order of localization time \cite{Goan00},
$1/\gamma_{\rm loc}^{\rm jump}=(D+D')/[\Gamma_d(\sqrt{D}+\sqrt{D'})^2]$.
In the parameter regime of Fig.\ \ref{fig:PC-Zeno}(c) ($\Gamma_d/\Omega=8$),
this time is still much smaller than the average time between state-changing
transitions (period of oscillations) due to $\Omega$, 
i.e., the mixing time \cite{Goan00}, 
$1/\gamma_{\rm mix}=\Gamma_d/(4\Omega^2)$. 
Hence, we can already see 
from  Fig.\ \ref{fig:PC-Zeno}(c) for $\Gamma_d/\Omega=8$ 
that very frequent repeated measurements would 
tend to localize the system.

\begin{figure}
\centerline{\psfig{file=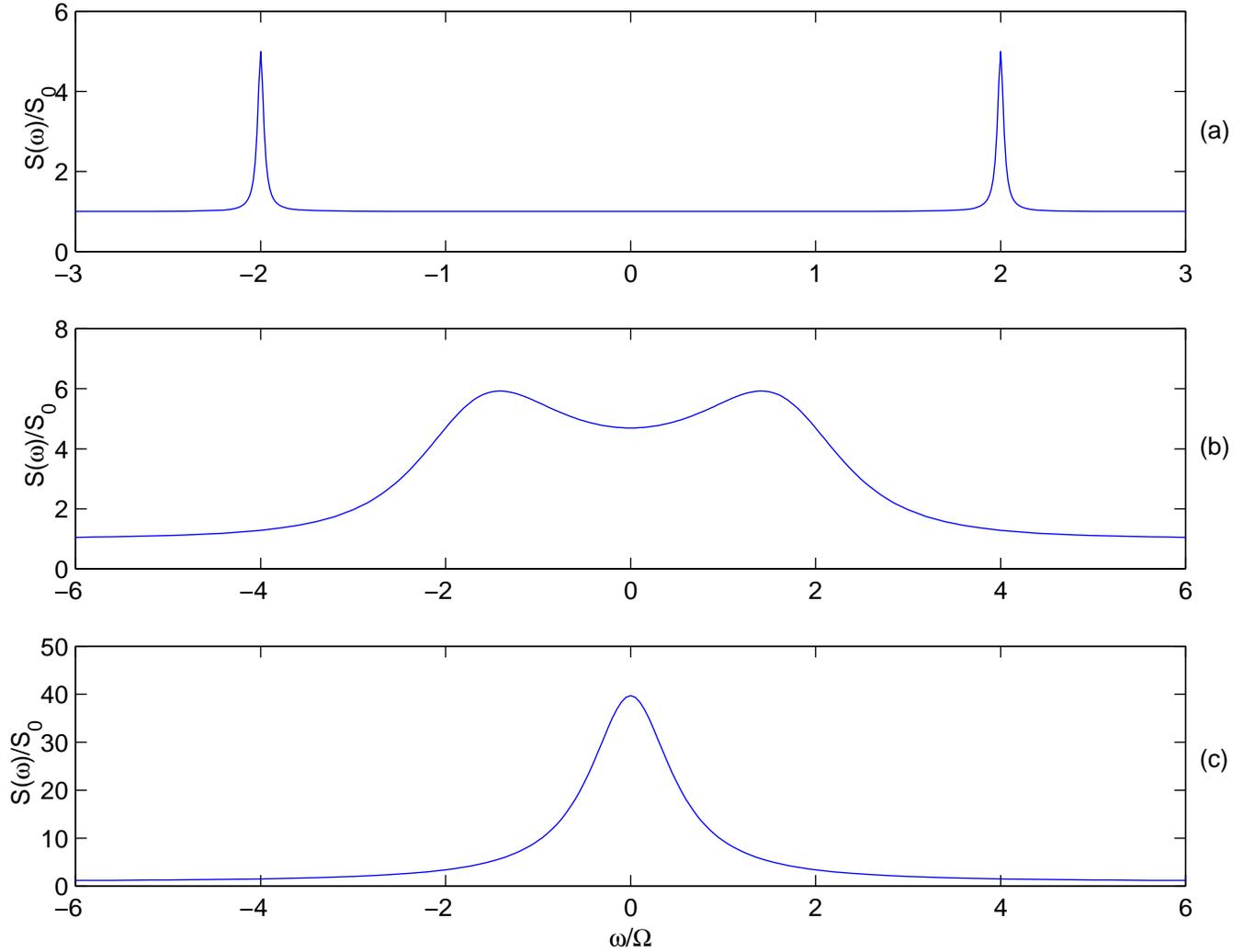,width=\linewidth,angle=0}}
\caption{A plot of the noise power spectrum of
the current, normalized by the shot noise level for different ratios
of (a)$(\Gamma_d/\Omega)$=0.04, (b) 2, (c) 8.
All the parameters are the same as the corresponding ones in
Fig.\ \ref{fig:PC-Zeno}.
For small $(\Gamma_d/\Omega)$ ratio, two sharp peaks appear in 
the noise power spectrum, as shown in (a).
In (b), a double peak structure is still visible,
indicating that coherent tunneling between the two qubit states
still exists.
In the classical, incoherent regime
$\Gamma_d \geq 4 \Omega$, only one single peak appears, as shown
in (c).} 
\label{fig:PC-noise}
\end{figure}

The ensemble average behavior of $z(t)$ is also shown 
in dashed line in
Fig.\ \ref{fig:PC-Zeno}.
If ${\cal E}=0$ and initially the electron is in
dot 1, from the solution of 
Eq.\ (\ref{rateEq}) 
, the probability $\rho_{aa}(t)=[1-z(t)]/2$
can be written as 
\begin{equation}
\rho_{aa}(t)
=\frac{1}{2}\left\{1+e^{-\Gamma_d t/2} 
\left[\cosh\left(\frac{\Omega_\Gamma}{2}t\right)+
\frac{\Gamma_d}{\Omega_\Gamma}\sinh\left(\frac{\Omega_\Gamma}{2}t\right)
\right]\right\},
\label{rhoaa}
\end{equation} 
where $\Omega_\Gamma=\sqrt{\Gamma_d^2-(4\Omega)^2}$. 
In the Appendix, 
the stationary noise power spectrum of
the current fluctuations through the PC barrier
is calculated for the case of ${\cal E}=0$ and 
the result can be written as \cite{Korotkov00}:
\begin{equation}
S(\omega)
=S_0+\frac{4  \Omega^2 (\Delta i)^2 \Gamma_d}
{(\omega^2-4\Omega^2)^2+\Gamma_d^2 \, \omega^2}.
\label{noisespectrum}
\end{equation}
where $S_0=2 \,e \, i_{\infty}= e^2\, \zeta\, (D'+D)$
represents the shot noise, $i_{\infty}=e\, \zeta\, (D'+D)/2$ is the
steady-state current and $\Delta i=e \, \zeta\, (D-D')$ 
represents the difference between the two average currents.
For $\Gamma_d < 4 \Omega$, $\rho_{aa}(t)$
shows the damped oscillatory behavior
in the immediate time regime 
(see dashed line in Fig.\ \ref{fig:PC-Zeno}(a) and (b)). 
In this case, the spectrum has a
double peak structure,
indicating that coherent tunneling is taking place
between the two qubit states. 
This is illustrated in Fig.\ \ref{fig:PC-noise}(a) and (b).
When $\Gamma_d \geq 4\Omega$,
$\rho_{aa}(t)$ does not oscillate but decays in time purely
exponentially, saturating at the
probability $1/2$. (see dashed line in Fig.\ \ref{fig:PC-Zeno}(c))
This corresponds to a classical,
incoherent behavior.
In this case, only a single peak, centering at $\omega=0$,
appears in the noise spectrum, 
as illustrated in Fig.\ \ref{fig:PC-noise}(c). 
The evolution of $z_c(t)$ in Fig.\ \ref{fig:PC-Zeno}(c), 
is one of the possible conditional evolutions
in this parameter regime ($\Gamma_d/\Omega=8$).
In this parameter regime $\Gamma_d \geq 4\Omega$,
the conditional evolution $z_c(t)$
behaves very close to a  
probabilistic jumping or random telegraph process.
After ensemble averaging over all possible realizations
of such conditional evolutions,  
one would then obtain the classical, incoherent behavior.

\subsection{Relative phase of the tunneling amplitudes}
\label{sec:phase}

The relative phase between the two complex tunneling
amplitudes produces effects on
both conditional and unconditional dynamics of the qubit.
In the following, we consider the case that $\zeta=1$ and  ${\cal E}=0$.
From Eq.\ (\ref{condBlochEq}),
after each jump the imaginary part of the product 
$({\cal T}^* {\cal X})$ seems to
cause an additional rotation around the $z$-axis in the Bloch sphere, 
but does not directly change the
population probability $z_c(t)$ of the qubit.
However, the actual conditional evolution of 
the Bloch sphere variables is complicated.
It is stochastic and nonlinear, and 
depends on the relative phase of the tunneling amplitudes
in a nontrivial way.
Nevertheless, after ensemble average, 
the imaginary part of $({\cal T}^* {\cal X})$ generates an effective
shift in the energy mismatch of the qubit states 
(see Eq.\ (\ref{rateEq})).

\begin{figure}
\centerline{\psfig{file=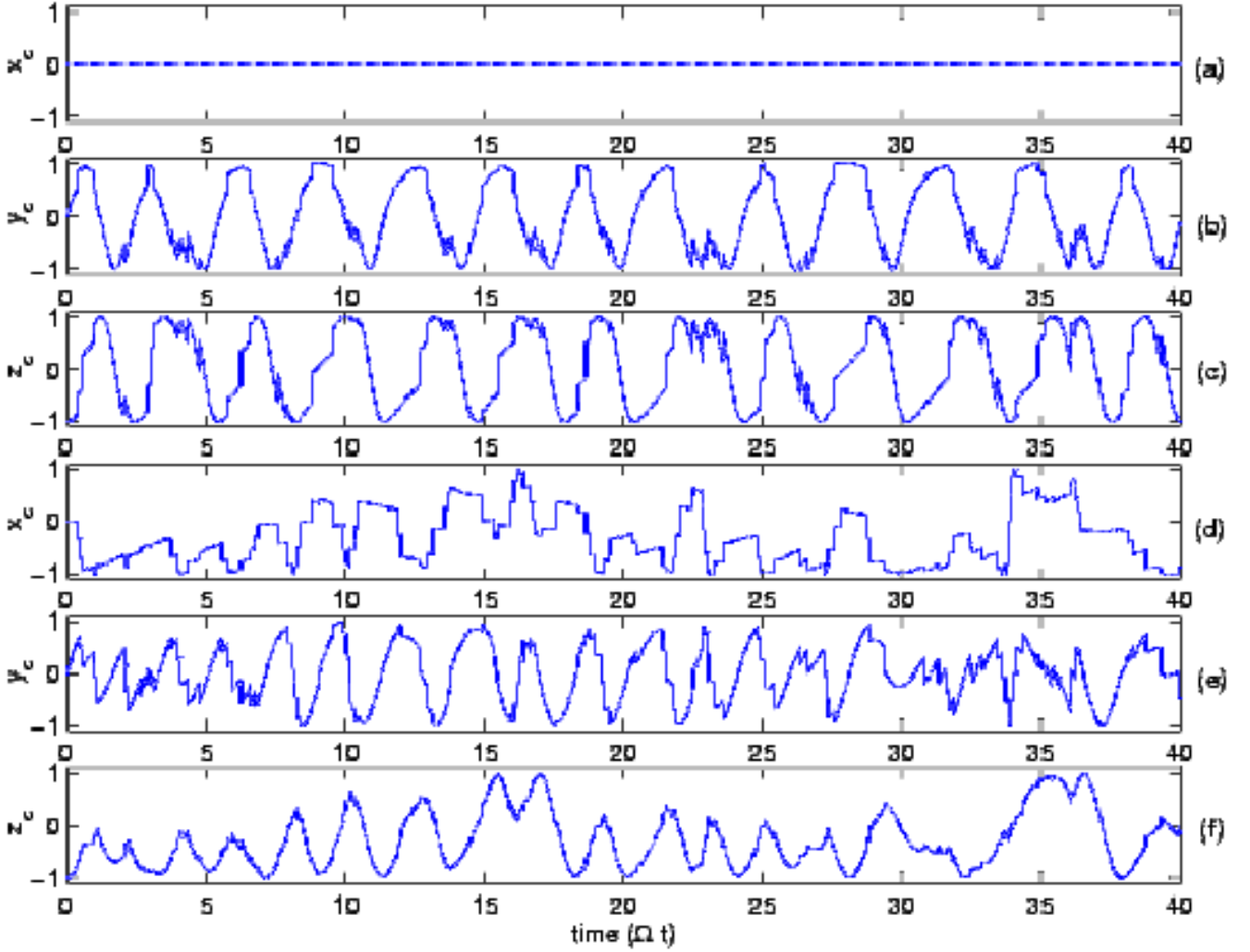,width=\linewidth,angle=0}}
\caption{Effect of relative phase on the qubit dynamics. 
The conditional evolutions of 
$x_c(t)$, $y_c(t)$, and $z_c(t)$ 
with the same initial condition
(the qubit being in $|a\rangle$)
and parameters ($\zeta=1$, 
${\cal E}=0$, $\theta=\pi$, $|{\cal T}|^2=4|{\cal X}|^2=4\Omega$),
but different relative phases are shown:
(a)--(c) for $\theta=\pi$, and (d)--(f) for
$\theta=\cos^{-1}(|{\cal X}|/| {\cal T}|)$.
The relative phase causes quite different evolutions for $x_c(t)$.} 
\label{fig:PC-phase}
\end{figure}

There are situations in which the effect of the 
relative phase of the tunneling amplitudes can be easily seen.
For $\zeta=1$ and  ${\cal E}=0$,
if the tunneling amplitudes are real, i.e., $\theta=\pi$,
and the initial condition $x_c(0)=0$, then the time evolution of $x_c(t)$, 
from Eq.\ (\ref{condBlochEq}), does not change and 
remains at the value $0$ at all times.
But if $\theta\neq\pi$ or $\sin\theta\neq 0$, 
the conditional evolution of $x_c(t)$ behaves rather differently.
It changes after the first detection (quantum jump) takes place.
Fig.\ \ref{fig:PC-phase} shows the evolutions of the Bloch variables 
$x_c(t)$, $y_c(t)$, $z_c(t)$ with the same initial condition
(the qubit being in $|a\rangle$)
and parameters but different relative phases:
$\theta=\pi$ for (a)--(c), and 
$\theta=\cos^{-1}(|{\cal X}|/| {\cal T}|)$ for (d)--(f).
We can clearly see quite different behaviors of $x_c(t)$ in these two
cases. The asymmetry of the electron population in $z_c(t)$,
due to effectively generated energy mismatch 
in the second case in Fig.\ \ref{fig:PC-phase}(f), 
can be roughly observed.   
The effect of the relative phase is small in the case of quantum diffusion.
As noted in Sec.\ \ref{sec:conditional}, 
in order for the quantum-diffusive equations
to be valid, we should
regard, to the order of magnitude, that 
$|\cos\theta|\sim O(1)$ and $|\sin\theta|\sim O(\epsilon)$.
This implies that in this case $\theta\approx \pi$. Hence 
the effect of the relative phase is small and
the conditional dynamics does not deviate
much from the case that the tunneling amplitudes are 
assumed to be real\cite{Korotkov99,Korotkov00,Korotkov00c}.

\subsection{Inefficient measurement and non-ideality}
\label{sec:inefficiency}
We have shown \cite{Goan00} that 
for $\zeta=1$, the conditional time evolution of the
qubit can be described by a ket state vector satisfying the 
stochastic Schr\"{o}dinger equation. It is then obvious that 
perfect detection or efficient measurement 
preserves state purity for a pure initial state. 
However, the inefficiency and non-ideality of
the detector spoils this picture.
The decrease in our knowledge of the qubit state
leads to partial decoherence for the qubit state.
We next find the partial decoherece rate introduced in this way.

The stochastic differential equations
in the form of It\^{o} calculus
\cite{Gardiner85,Oksendal92}
have the advantage that it is easy to see that 
the ensemble average of the conditional
equations over the random process
$\xi(t)$ leads to the unconditional equations.
However, it is not a natural physical choice. For example, 
for $\zeta=1$, the term
$-\Gamma_d \rho_{ab}(t)$ in Eq.\ (\ref{diffrateEqc}) does not really
cause decoherence of the conditional qubit density matrix. 
It simply compensates the noise term due to the
definition of derivative in It\^{o} calculus.
Hence, in this case the conditional evolution of $\rho_{ab}(t)$
does not really decrease in time exponentially. 
To find the partial decoherence rate
generated by inefficiency $\zeta<1$, we transform 
Eq.\ (\ref{diffrateEqc}) into the form of Stratonovich
calculus \cite{Gardiner85,Oksendal92}. 
We then obtain for $\theta=\pi$:
\begin{equation}
\dot\rho_{ab}(t) =  i{\cal E}
\rho_{ab}(t)+i\Omega[\rho_{aa}(t)
-\rho_{bb}(t)]
-[\rho_{bb}(t)-\rho_{aa}(t)]\,
\frac{\sqrt{2\Gamma_d}}{e|{\cal T}|}\,[i(t)-i_0]
\, \rho_{ab}(t)-(1-\zeta)\Gamma_d\,\rho_{ab}(t),
\label{diffStratonovich}
\end{equation}
where 
$i(t)-i_0=e|{\cal T}|
\{\zeta \,  \sqrt{2\Gamma_d}[1-2\rho_{aa}(t)]+\sqrt{\zeta\, } \,
\xi(t)\}$.
Here we have used the following relations:
the conditional current $i(t)=e\, \delta N(t)/\delta t$ 
with $\delta N(t)$ given
by Eq.\ (\ref{deltaN}) and the average current  
$i_0=e\, \zeta\, (D+D')/2$, where
$D=|{\cal T}|^2$ and $D'=|{\cal T}|^2-2|{\cal T}||{\cal X}|$
in the quantum-diffusive limit.
In this form, Eq.\ (\ref{diffStratonovich}) elegantly 
shows how the qubit
density matrix is conditioned on the measured current.
We find that the last term in Eq.\ (\ref{diffStratonovich})
is responsible for decoherence.
In other words, the partial decoherence rate for 
an individual realization of inefficient measurements is 
$(1-\zeta)\Gamma_d$. For a perfect detector 
$\zeta=1$, this decoherence rate vanishes and 
the conditional $\rho_{ab}(t)$, as expected, 
does not decay exponentially in time.
Similar conclusion could be drawn from Eq.\ (\ref{condBlochEq}) for the
quantum-jump case. For $\theta=\pi$, the off-diagonal variables
$x_c(t)$ and $y_c(t)$ seem to decrease in time with the rate 
$(1-\zeta)\Gamma_d$.

\begin{figure}
\centerline{\psfig{file=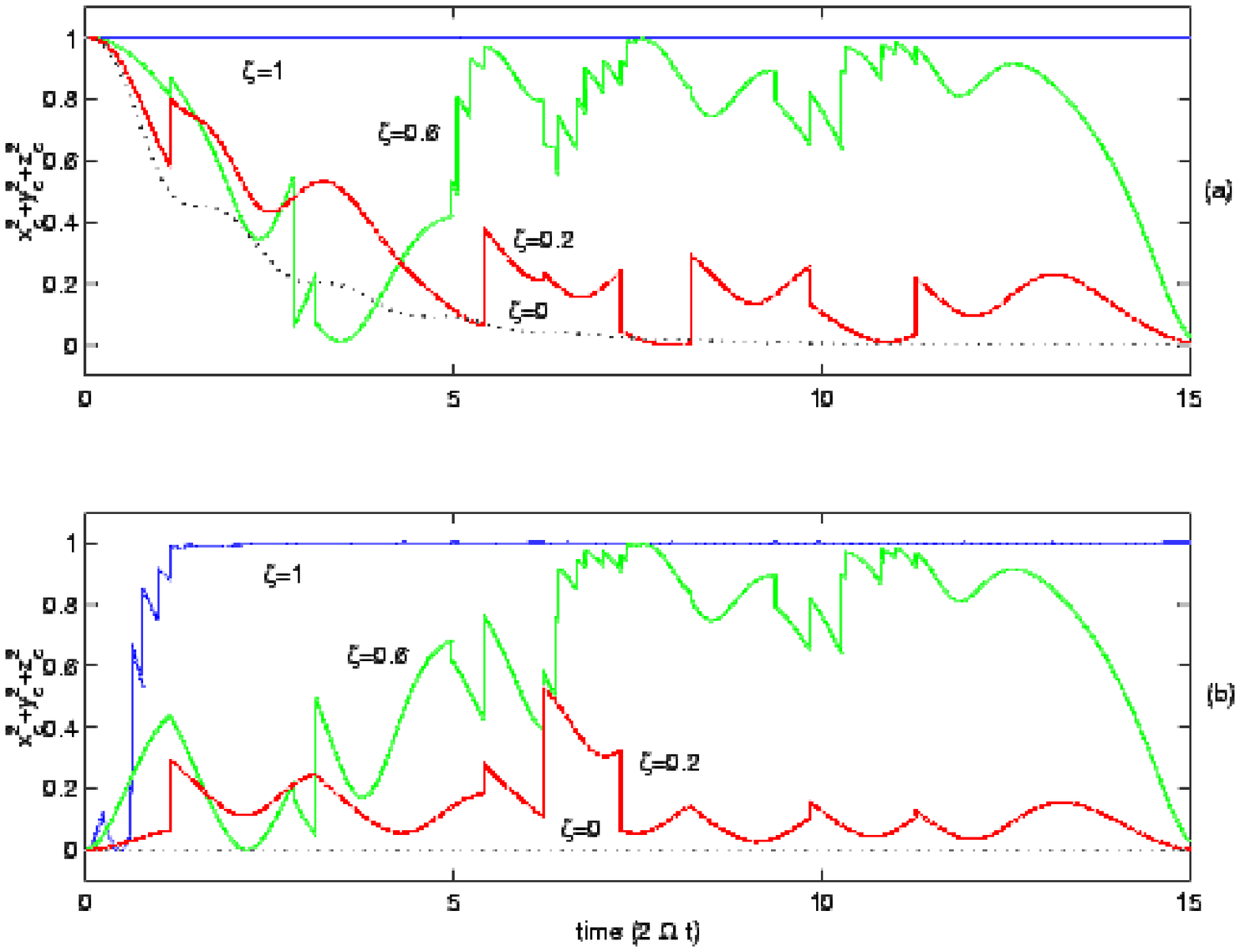,width=\linewidth,angle=0}}
\caption{Effect of inefficiency on the state purity.
The quantum-jump, conditional evolution of the 
purity $P_c(t)$ for different inefficiencies,
$\zeta=1$, $0.6$, $0.2$ (in solid line), and $0$ (in dotted line) are plotted 
in (a) for a initial qubit state being in a pure state $|a\rangle$,
(b) for a maximally mixed initial state.
The other parameters are:
${\cal E}=0$, $\theta=\pi$, $|{\cal T}|^2=4|{\cal X}|^2=4\Omega$.
The purity-preserving conditional evolution for a pure initial state,
and gradual purification for a non-pure initial state for $\zeta=1$
are illustrated. However,
the complete purification of the qubit state can not
be achieved for $\zeta<1$.} 
\label{fig:PC-purity}
\end{figure}

In Bloch sphere variable representation, we can use the quantity
$P_c(t)=x_c^2(t)+y_c^2(t)+z_c^2(t)$ as a measure of the purity of the
qubit state, or equivalently as a measure of how much information the
conditional measurement record gives about the qubit state.
If the conditional state of the qubit is a pure state then
$P_c(t)=1$; if it is a maximally incoherent mixed state then $P_c(t)=0$. 
We plot in Fig.\ \ref{fig:PC-purity}
the quantum-jump, conditional evolution of the 
purity $P_c(t)$ for different inefficiencies,
$\zeta=1$, $0.6$, $0.2$ (in solid line), and $0$ (in dotted line).
Fig.\ \ref{fig:PC-purity}(a) is 
for a initial qubit state being in a pure state $|a\rangle$,
while Fig.\ \ref{fig:PC-purity}(b) is for a maximally mixed initial state.   
We can see from Fig.\ \ref{fig:PC-purity}(a) that the purity $P_c(t)=1$ at all
times for $\zeta=1$, while it hardly or 
not at all reaches $1$ 
for almost all time for $\zeta<1$.
This means that partial information about the changes of the
qubit state is lost irretrievably in inefficient measurements.
In addition, roughly speaking, the overall behavior of $P_c(t)$ 
decreases with decreasing $\zeta$.
This indicates that after being averaged over a long period of time,
$\langle P_c(t)\rangle_t$ would also decrease with decreasing $\zeta$.  
For $\zeta=0$, the evolution of $P(t)$ becomes smooth and tends toward
the value zero (the maximally mixed steady state).
For a non-pure initial state (see Fig.\ \ref{fig:PC-purity}(b)),
the qubit state is eventually 
collapsed towards a pure state 
and then remains in a pure state for $\zeta=1$.
But the complete purification of the qubit state cannot
be achieved for $\zeta<1$. 
As in Fig.\ \ref{fig:PC-jump-diff}(a)--(d),
the same randomness has been applied to 
generate the quantum-jump,
conditional evolution 
in Fig.\ \ref{fig:PC-purity}(a) and (b).
Note that 
the only difference between evolution in Fig.\ \ref{fig:PC-purity}(a) 
and the corresponding one in Fig.\ \ref{fig:PC-purity}(b) is the
different initial states.
So when the qubit density matrix 
in Fig.\ \ref{fig:PC-purity}(b)
gradually evolves into the same state as in  
Fig.\ \ref{fig:PC-purity}(a), the corresponding $P_c(t)$ in 
Fig.\ \ref{fig:PC-purity}(b) would then follow the same evolution
as in Fig.\ \ref{fig:PC-purity}(a). 
This behavior can be observed in Fig.\ \ref{fig:PC-purity}.
The purity-preserving conditional evolution for a pure initial state,
and gradual purification for a non-pure initial state for an ideal detector
have been discussed in
Refs.\ \cite{Korotkov99,Korotkov99b,Korotkov00,Korotkov00b}
in the quantum-diffusive limit.

The non-ideality of the PC detector is modeled in 
Refs.\ \cite{Korotkov99,Korotkov99b,Korotkov00,Korotkov00b}
by another ideal detector ``in parallel'' to the original one 
but with inaccessible output.
We can add, as in 
Refs. \cite{Korotkov99,Korotkov99b,Korotkov00,Korotkov00b},
an extra term, $-\gamma_d\rho_{ab}(t)$, to 
Eq.\ (\ref{diffStratonovich})  
to account for the ``non-ideality'' of the detector.
The ideal factor $\eta$ introduced 
there \cite{Korotkov99,Korotkov99b,Korotkov00,Korotkov00b}
can be modified to take account of inefficient measurement 
discussed here. We find
\begin{equation}
\eta=1-\frac{\Gamma}{\Gamma_{tot}}=\frac{\zeta \,  \Gamma_d}{\Gamma_d+\gamma_d},
\end{equation}
where 
$\Gamma=(1-\zeta)\Gamma_d+\gamma_d$ and $\Gamma_{tot}=\Gamma_d+\gamma_d$.
For $\gamma_d=0$, we have $\eta=\zeta$. 
In Ref.\ \cite{Korotkov00c}, inefficient measurement is discussed
in terms of insufficiently small readout period.
As a result, the information about the tunneling times
of the electrons passing through the PC barrier is partially lost.

\section{Conclusion}
\label{sec:conclusion}

We have obtained 
the quantum-jump and quantum-diffusive, 
conditional master equations 
for the case of inefficient measurements.
These conditional master equations describe the 
random evolution of the measured qubit density matrix, which both
conditions and is 
conditioned on, a particular realization of the measured current.
We have analyzed the conditional qubit dynamics in detail and
illustrated the conditional evolution by numerical simulations.
Specifically, the conditional qubit dynamics evolving
from quantum jumps to quantum diffusion has been presented. 
Furthermore, we have described
the quantum Zeno effect 
in terms of the quantum-jump, conditional
dynamics.
We have also discussed
the effect of inefficient measurement and the influence of
relative phase between the two
tunneling amplitudes
on the qubit dynamics.

\section*{Acknowledgment}
\label{sec:Acknow}

H.S.G. is grateful for useful discussions with A.~N.~Korotkov,
G.~Sch\"on, D.~Loss, Y.~Hirayama, J.~S.~Tsai and G.~P.~Berman.
H.S.G. would like to thank H.~B.~Sun and H.~M.~Wiseman
for their assistance and discussions in the early stage of this work.

\appendix
\section{Calculation of the noise power spectrum of
the current fluctuations}
\label{sec:noisespectrum}

In this Appendix, we calculate the stationary noise power spectrum of
the current fluctuations 
through the PC when there is the possibility of coherent
tunneling between the two qubit states.
Usually one can calculate this noise power spectrum using the unconditional,
determistic master equation approach, which gives only the 
average characteristics. 
We, however, calculate it through the
stochastic formalism presented here. 
The fluctuations in the observed current, $i(t)$, are quantified by
the two-time correlation function:
\begin{equation}
G(\tau )= E[i(t+\tau)i(t)]-E[i(t+\tau)]\, E[i(t)]
\end{equation}
The noise power spectrum of the current is then given by 
\begin{equation}
S(\omega )=2 \int_{-\infty}^{\infty }d\tau \, G(\tau )\, e^{-i \omega \tau}.
\label{Sw}
\end{equation}
The ensemble expectation values of the two-time correlation function
for the current in the case of quantum diffusion has been
calculated in Ref.\ \cite{Korotkov00}.
Here we will present the quantum-jump case.
The current in this case is given by Eq.\ (\ref{current}).
We will follow closely the calculation in the Appendix of 
Ref.\ \cite{Wiseman93} to calculate the two-time correlation function,
$E[dN_c(t+\tau)dN(t)]$. 
First we consider the case when $\tau\gg dt>0$,
where $dt$ is the minimum time step considered. 
Since $dN(t)$ is a classical point process, it is either zero or one.
As a result, $E[dN_c(t+\tau)dN(t)]$ is non-vanishing only if there is
electron-tunneling event inside each of these two infinitesimal time
intervals, $[t,t+dt]$ and $[t+\tau,t+\tau+dt]$.
Hence, we can write
\begin{equation}
E[dN_c(t+\tau)dN(t)]={\rm Prob}[dN(t)=1]\, E[dN_c(t+\tau)|_{dN(t)=1}],
\label{dNdN}
\end{equation}
where the subscript to the vertical line is the condition for which
the subscript on $dN_c(t+\tau)$ exists.  
From Eqs.\ (\ref{ineffdNav}) and (\ref{measuretilderho1}), we have
${\rm Prob}[dN(t)=1]=\zeta \, {\rm Tr}[\tilde\rho_{1}(t+dt)]$ and 
$E[dN_c(t+\tau)|_{dN(t)=1}]=\zeta\, {\rm Tr}
\{{\cal J}[{\cal T}+{\cal X} n_1] 
E[\rho_{1c}(t+\tau)|_{dN(t)=1}]\} $. 
Using the fact that $E[\rho_c(t)]=\rho(t)$ and 
Eqs.\ (\ref{Liouvillian}) and (\ref{measuretilderho1}),
we can write
\begin{eqnarray}
E[\rho_{1c}(t+\tau)|_{dN(t)=1}]
&=&e^{{\cal L}(\tau-dt)} 
\tilde\rho_{1}(t+dt)/{\rm Tr}[\tilde\rho_{1}(t+dt)] 
\nonumber 
\\ 
&=&\zeta \, e^{{\cal L}(\tau-dt)}
\{{\cal J}[{\cal T}+{\cal X} n_1]\rho(t) dt \}
/{\rm Tr}[\tilde\rho_{1}(t+dt)].
\label{rhoL}
\end{eqnarray}
Hence, to leading order in $dt$, we obtain for $\tau>0$:
\begin{equation}
E[dN_c(t+\tau)dN(t)]=\zeta^2 dt^2 
{\rm Tr}
\left[{\cal J}[{\cal T}+{\cal X} n_1] 
e^{{\cal L}\tau}\{{\cal J}[{\cal T}+{\cal X}
n_1]\rho(t)\}\right].
\label{taut}
\end{equation}
For $\tau=0$, we have, from Eq.\ (\ref{pointprocess}), that:
\begin{equation}
E[dN(t)dN(t)]=E[dN(t)]=\zeta\, [D+(D'-D)\langle n_1\rangle(t)] dt.
\label{tau0}
\end{equation}
For short times, this term dominates and we may regard $dN(t)/dt$
as $\delta$-correlated noise for a suitably defined $\delta$ function.
Thus the current-current two-time correlation function for 
$\tau\geq 0$ can be written as: 
\begin{eqnarray}
E[i(t+\tau)i(t)]
&=&E[\frac{dN_c(t+\tau)}{dt}\frac{dN(t)}{dt}]
\nonumber \\
&=& e^2\, \zeta\, \{D+(D'-D){\rm Tr}[ n_1 \rho(t)]\}\, \delta(\tau)
+\zeta^2\, {\rm Tr}
\left[{\cal J}[{\cal T}+{\cal X} n_1] 
e^{{\cal L}\tau}\{{\cal J}[{\cal T}+{\cal X}
n_1]\rho(t)\}\right].
\label{ii}
\end{eqnarray}
In this form, we have related the ensemble averages of classical
random variable to the
quantum averages with respect to the qubit density matrix. 
The case $\tau\leq 0$ is covered by the fact that the 
current-current two-time correlation function or $G(\tau)$ is
symmetric in $\tau$, i.e., $G(\tau)=G(-\tau)$.

Next we calculate steady-state $G(\tau)$ and $S(\omega)$.
We can simplify Eq.\ (\ref{ii})
using the following identities
for an arbitrary operator $B$: 
${\rm Tr}\left[ {{\cal J}[n_{1}]B}\right] \equiv {\rm Tr}[n_{1}B]$, 
${\rm Tr}[e^{{\cal L}\tau}B]={\rm Tr}[B]$, 
and ${\rm Tr}[Be^{{\cal L\tau }}\rho_{\infty}]={\rm Tr}[B\rho _{\infty }]$, 
where the $\infty$ subscript indicates that the system is at
the steady state and the steady-state density matrix $\rho_{\infty}$ is  
a maximally mixed state. 
Hence we obtain the steady-state $G(\tau)$ for $\tau\geq 0$ as:
\begin{equation}
G(\tau )=e \, i_{\infty }\delta ({\tau })+
e^{2}\, \zeta^2\, (D'-D)^2
\left\{ {\rm Tr}\left[ n_1 e^{{\cal L}\tau }
[n_1 \rho_{\infty}\right] -{\rm Tr}[n_1\rho_{\infty}]^2\right\},
\label{Gt}
\end{equation}
where the steady-state average current $i_{\infty}= e \, \zeta\, (D+D')/2$. 
The first term in Eq.\ (\ref{Gt}) represents the shot noise component.
It is easy to evaluate Eq.\ (\ref{Gt}) analytically for ${\cal E}=0$ case.
The case for the asymmetric qubit, ${\cal E}\neq 0$, can be calculated
numerically.
Evaluating Eq.\ (\ref{Gt}) for ${\cal E}=0$, we find
\begin{equation}
G(\tau )=e \, i_{\infty }\delta ({\tau })
+\frac{(\Delta i)^2}{4}
\left( \frac{\mu_{+}e^{\mu _{-}\tau }-\mu_{-}e^{\mu_{+}\tau }}
{\mu_+-\mu_-}\right) ,
\end{equation}
where $\mu_{\pm }=-(\Gamma_d/2)\pm \sqrt{(\Gamma_d/2)^2-4 \Omega^{2}}$,
and we have represented $\Delta i=e \, \zeta\, (D-D')$
as the difference between the two average currents.
After Fourier transform following from 
Eq.\ (\ref{Sw}),
the power spectrum of the noise 
is then obtained as the expression of Eq.\ (\ref{noisespectrum}).
Note that from Eq.\ (\ref{noisespectrum}),
the noise spectrum at $\omega=2\Omega$  
for $\theta=\pi$, i.e., real tunneling amplitudes, 
can be written as:
\begin{equation}
\frac{S(2\Omega)-S_0}{S_0}=2\, \zeta \,
\frac{(\sqrt{D}+\sqrt{D'})^2}{(D+D')}
\label{Swratio}
\end{equation}
where $S_0=2\, e\, i_{\infty}= e^2\, \zeta\, (D'+D)$ represents the shot noise.
In obtaining Eq.\ (\ref{Swratio}),
we have used the relation $\Gamma_d=(\sqrt{D}-\sqrt{D'})^2/2$ 
for the case of real tunneling amplitudes.
In the quantum-diffusive limit $|{\cal T}|\gg |{\cal X}|$
or $(D+D')\gg(D-D')$,
this ratio 
 $[S(2\Omega)-S_0]/{S_0}\to 4 \zeta$, independent \cite{Korotkov00} 
of the values
of $\Omega$ and $\Gamma_d$. 
These results for $\zeta=1$ and in the limit of quantum diffusion
are consistent with those derived  
in Ref.\ \cite{Korotkov00}
using both the unconditional master equation approach and conditional
stochastic formalism with 
white noise current fluctuations for an ideal detector.


\newpage

\begin{references}
\bibitem{temperature} The finite-temperature unconditional 
master equation for the CQD system,
taking into account the effect of thermal PC
reservoirs under the weak system-environment coupling
and Markovian approximations, is obtained in Ref.\ \cite{Goan00}.
\end{references}
\end{document}